\documentclass[prb,aps, twocolumn, amsmath,amssymb,superscriptaddress]{revtex4}
\usepackage{graphicx} 
\usepackage{float}
\usepackage{amsmath}
\usepackage{amssymb}
\usepackage{amsfonts}
\usepackage{euscript}
\usepackage{enumerate}
\usepackage{hhline}
\usepackage{tabularx}

\usepackage{booktabs}

\usepackage[dvipsnames]{xcolor} 

\newcommand{\YS}[1]{{\color{blue}#1}}

\usepackage{dcolumn}
\usepackage{bm}
\usepackage{soul}

\makeatletter
\renewcommand{\@biblabel}[1]{#1. }
\renewcommand{\@dotsep}{500}
\renewcommand{\@pnumwidth}{0em}
\renewcommand{\l@figure}[2]{
\@dottedtocline{1}{1.5em}{2em}{Figure #1}{}\vspace{15pt}}

\usepackage[colorlinks,linkcolor=red,anchorcolor=blue,citecolor=blue]{hyperref}

\begin{document}

\title{Freespace twistronics for optical supertopologies}

\author{Vasu Dev}
\affiliation{Centre for Disruptive Photonic Technologies, School of Physical and Mathematical Sciences, Nanyang Technological University, Singapore 637371, Republic of Singapore}

\author{Yijie Shen}
\email{yijie.shen@ntu.edu.sg}
\affiliation{Centre for Disruptive Photonic Technologies, School of Physical and Mathematical Sciences, Nanyang Technological University, Singapore 637371, Republic of Singapore}
\affiliation{School of Electrical and Electronic Engineering, Nanyang Technological University, Singapore 639798, Singapore}

\date{\today}

\begin{abstract}
\noindent \textbf{Twistronics, the study of moir\'e superlattices of twisted bilayer 2D materials creating nontrivial physical effects, has recently revolutionized diverse subjects from materials to optoelectronics, nanophotonics, and beyond. 
Here, breaking the reliance on materials, we present twistronics in higher-dimensional free space, where the twisted lattice is not a layer of 2D material but a 3D propagating light field with topological textures. Moiré structured light with a twist angle can generate a rich set of high-dimensional topologies, including skyrmionium bags, skyrmion bag superlattices, skyrmion clusters, and optical quasicrystals, with controllable symmetry. Many of these textures have not been reported before. Importantly, in contrast to prior moiré superlattices, our freespace optical moiré textures maintain their topologies over a long propagation distances, showing nondiffractive behavior and robustness against perturbations and obstacles. Our work unlocks higher dimensions to manipulate moir\'e photonics with high-capacity topologies to address modern challenges of robust information transfer and encryption.}
\end{abstract}

\maketitle

\textit{Twistronics} -- the study of the evolution of physical properties in two-dimensional materials induced by a controlled twist between adjacent layers. The interference of two periodic lattices with a relative twist induces a moiré pattern, which has transited from everyday textiles to the scientific frontier~\cite{carr2017twistronics}. Remarkably, the moir\'e superlattice patterns on the twisted  two atomically thin layers can generate entirely novel physical properties with ``$1+1\gg2$'' scheme, e.g., the superconductivity~\cite{cao2018unconventional,park2021tunable}, correlated insulator behaviour~\cite{cao2018correlated}, tunable spin-polarized phases~\cite{cao2020tunable}, flat-band and quantum geometry~\cite{torma2022superconductivity} in twisted bilayer and multilayer graphenes. Moreover, twistronics has been applied to a variety of advanced materials, where specific twist angles known as `magic' angles can radically alter the electronic and photonic properties of materials~\cite{du2023moire}. For instance, the excitonic devices with twisted van der Waals heterostructures~\cite{ciarrocchi2022excitonic}, photonic hyperbolic polaritons excited on twisted $\alpha$-MoO$_3$~\cite{hu2020topological}, optical Hall effect enabled by spiral multilayer WS$_2$~\cite{ji2024opto}. Twisted photonic crystals can be used to localize light fields~\cite{wang2020localization} and enhance reconfigurable and singular nanolasers~\cite{luan2023reconfigurable,ouyang2024singular}. Recently, twistronics was extended to hydrodynamics to control energy localization in fluid~\cite{xu2024hydrodynamic}, and also to surface plasmon plariton to create nontrivial topological textures of evanescent waves on metal surface~\cite{schwab2025skyrmion}.

However, the existing concept of twistronics relies on material platforms, where the related physical phenomena occur primarily at low-dimensional surfaces. 
Furthermore, the resulting moiré phenomena are typically confined to near-field interactions or surface-bound excitations, leaving their extension into freely propagating wave systems largely unexplored. These constraints highlight the need for a framework that can realize moiré physics and topological interactions in higher-dimensional, reconfigurable physical systems.

The investigation of novel topological quasiparticles, including and beyond skyrmions, has become a central pursuit in modern physics aimed at achieving robust and reconfigurable information carriers. Recent advances have revealed a wide range of such excitations across diverse physical systems, including antiferromagnetic merons~\cite{jani2021antiferromagnetic}, magnetic skyrmion bundles~\cite{tang2021magnetic}, hopfions~\cite{zheng2023hopfion}, torons in liquid crystals~\cite{wu2022hopfions}, optical spin textures~\cite{wu2025photonic}, optical skyrmion and meron lattices~\cite{tsesses2018optical,lei2021photonic,marco2024periodic,marco2024propagation,teng2025topological,wu2025optical}, hopfion crystals~\cite{lin2025space}, and other complex topologies observed in both light and matter systems~\cite{shen2024optical,bogdanov2020physical}.

Here we extend the principles of twistronics into the emerging domain of topological light waves, where structured light fields carry topological quasiparticle textures and propagate in free space~\cite{shen2025free}. Freespace twistronics provides a unified framework for generating a broad family of optical \textit{supertopologies}. 
These supertopologies include skyrmionium bags, skyrmion bags, skyrmion clusters, meron-antimeron cluster and topological quasicrystals that propagate nondiffractively and remain stable against perturbations, representing a new class of topological light states beyond conventional material systems. 

Important features of the supertopologies: (1) Supertopologies are generated by twisting two volumes of topological optical lattices, giving rise to complex moiré interference fields with novel topologies.
The same framework can also reproduce recently discovered configurations such as topological quasicrystals~\cite{tsesses2025four,putley2025mixing,lin2025photonic} and skyrmion bags~\cite{schwab2025skyrmion} in plasmonic systems. (2) These supertopologies propagate in free space over extended distances without diffraction, maintaining both their geometric scale and topological structure. (3) Supertopologies are intrinsically resilient to strong perturbations and exhibit self-healing in topology upon partial obstruction. Therefore, our results establish freespace twistronics as a foundational platform for investigating and engineering topological properties of light.

\begin{center}
	\begin{figure*}
		\begin{center}
			\includegraphics[width=\linewidth]{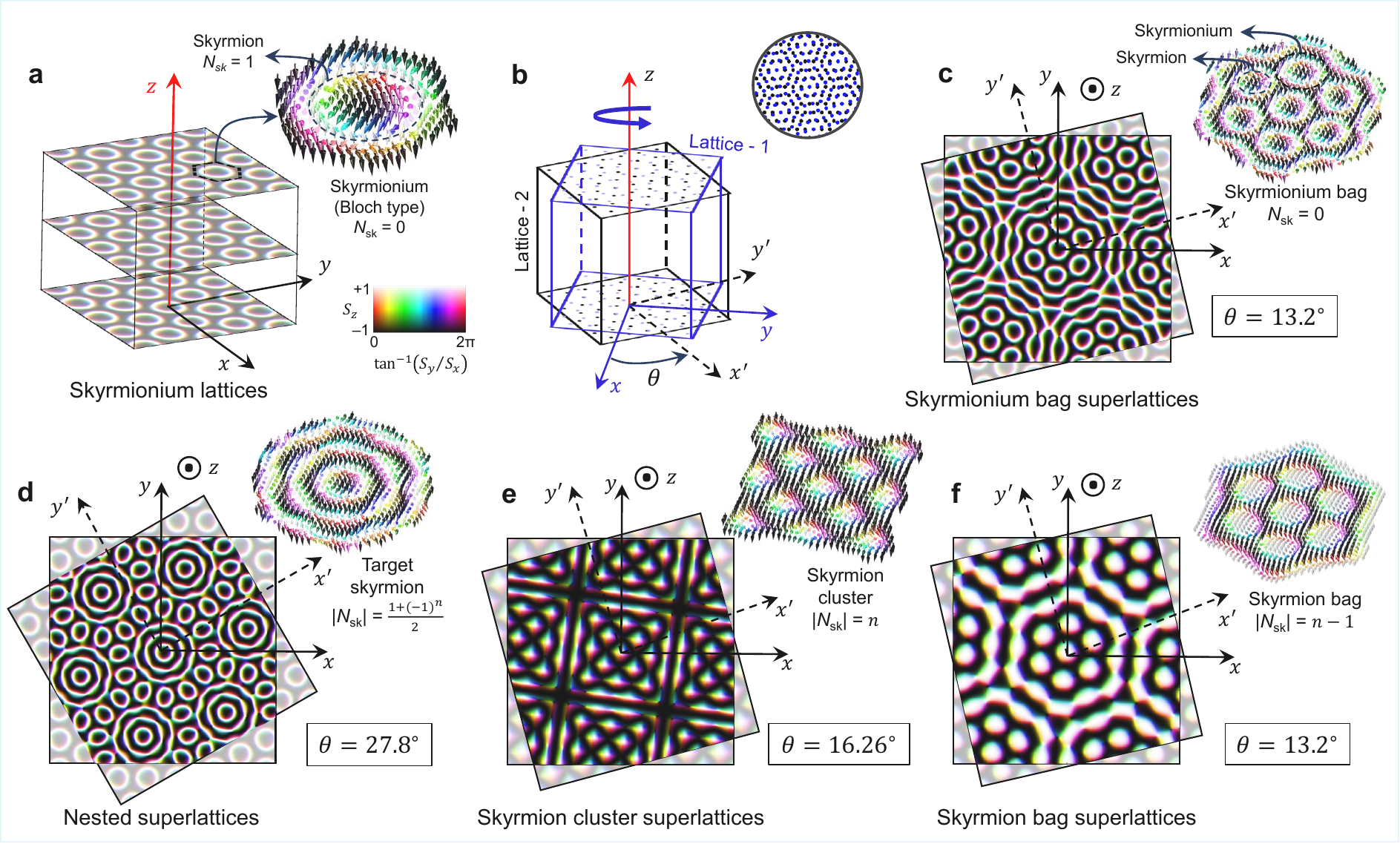}
			\caption{\textbf{Concept of freespace twistronics and supertopologies.}  (a) A basic freespace topological optical lattice field, which hosts C$_6$ symmetric skyrmionium lattices at each $x$-$y$ plane and propagate along $z$-axis with nondiffraction. The inset highlights the skyrmionium texture as an unit cell at a transverse plane, where the inner skyrmion with unit topology is marked by dashed line. (b) Schematic of the mori\'e pattern formed by superimposing two topological optical lattice fields (blue and black, where the dots represent centers of unit cells) with a twisted angle $\theta$ to $z$-axis. The inset shows a schematic mori\'e superlattice pattern at a transverse plane. (c) The resultant mori\'e structured light shows skyrmionium bag superlattices when $\theta=13.2^{\circ}$. The inset shows a skyrmionium bag, where multiple elementary skyrmioniums are surrounded by a bag skyrmioniums. (d) Nested superlattices with target-skyrmion-skyrmionium mixed topology obtained when $\theta=27.8^{\circ}$. The inset highlights a target skyrmion composed by multiple radially nested skyrmions with number of $n>2$. (e) Skyrmion cluster superlattices in twisted bi-volume C$_4$ skyrmion lattices when $\theta=16.26^{\circ}$. The inset shows a skyrmion cluster texture composed by $n$ skyrmions with total skyrmion number of $|N_\text{sk}|=n$. (f) Skyrmion bag superlattices in twisted bi-volume C$_6$ skyrmion lattices, when $\theta=13.2^{\circ}$. The insert shows a skyrmion bag with $n$ elementary skyrmions surrounded by a bag skyrmion with total skyrmion number of $|N_\text{sk}|=n-1$.}
			\label{F1}
		\end{center}
	\end{figure*}
\end{center}

\section{freespace twistronics}
Freespace twistronics establishes high-dimensional topological optical lattices as a new platform for moiré engineering, transcending the limits of two-dimensional materials and plasmonic media. 
In contrast to conventional twistronics, where moiré effects arise from electronic coupling between stacked layers, the optical version relies purely on field superposition and is therefore free from material constraints. This freedom enables access to higher-dimensional and dynamically reconfigurable moiré geometries that are challenging to realize in condensed matter systems.

We first develop a method to generate topological optical lattices with high phase stability and accuracy, and then apply freespace twistronics to obtain nontrivial high-dimensional topologies. The lattices are generated through the principle of $N$-wave interference, in which two structured light fields with binary and helical phase distributions and orthogonal polarisations are superimposed. In the Fourier domain, such lattices can be expressed as an array of $N$ discrete point sources, which is experimentally challenging to realize. Therefore, to simplify, we approximate intensity of each point source with a Gaussian distribution. Therefore, the topological lattice in the Fourier domain can be expressed as:

\begin{equation}
    U_F = \sum_{j=1}^{\mathrm{N}} A_1~e^{-\frac{(x-\alpha_j)^2 + (y-\beta_j)^2}{\omega^2_{0}}} \mathbf{\hat{e}_R} + e^{i\theta_0} A_2~e^{-\frac{(x-\alpha_j)^2 + (y-\beta_j)^2}{\omega^2_{0}}}~e^{i\phi_{j}} \mathbf{\hat{e}_L}.
\end{equation}

Here $A_1$ and $A_2$ denote the amplitudes of the right- and left-circularly polarised light fields, $\mathbf{\hat{e}_R}$ and $\mathbf{\hat{e}_L}$. The coordinates of the $j$th Gaussian beamlet are given by $(\alpha_j, \beta_j) = \gamma (\cos \delta_j, \sin \delta_j)$, where $\gamma = d / \sqrt{1 - \cos(2\pi / N)}$ and $\delta_j = \pi (2j - 1) / N$ with $j = 1, 2, \dots, N$. The parameter $d$ represents the centre-to-centre separation between neighbouring Gaussian beamlets with the same beam waist $\omega_0$. The discrete helical phase of each beamlet in left circularly polarized field is defined as $\phi_j = m [2\pi (j - 1) / N]$, where $m$ is the topological charge, and $\theta_0$ is the global phase difference between the right- and left-circularly polarised fields. Following this formulation, we generate topological optical lattices exhibiting three characteristic symmetries: hexagonal C$_6$ skyrmionium lattices ($N = 6$), square C$_4$ skyrmion lattices ($N = 4$), and triangular C$_3$ meron–antimeron lattices ($N = 3$) {\color{blue}(see the Extended Data Fig. 1).}

Twisting two such topological lattices with a controlled relative angle gives rise to a freespace moiré field, where the superposition of the individual textures produces entirely new optical structures with novel topologies. The twist couples the intrinsic symmetry of each lattice to an additional spatial degree of freedom, leading to the formation of moiré superlattices with modified periodicity and symmetry. By varying the twist angle and the symmetry of the constituent lattices, we can engineer a broad range of high dimensional topological configurations that do not exist in the individual lattices. This tunability forms the basis of freespace twistronics, allowing direct optical access to moiré physics and emergent topology without the need for material interfaces. As the twist angle increases, the resultant field evolves from regular lattice geometry to complex quasiperiodic and clustered geometries. This continuous transformation marks the onset of high dimensional topological organization in free space, as shown in Fig.~\ref{F1}. For instance, a C$_6$ skyrmionium lattice beam is shown in Fig.~\ref{F1}(a), where the polarisation Stokes vectors form a periodic array of skyrmioniums with sixfold rotational symmetry. Each unit cell is a hexagonal skyrmionium composed of two radially nested skyrmions with opposite polarity, giving a total skyrmion number of $N_\text{sk}=0$ (see inset). These topological lattices propagate along the $z$ axis without diffraction, maintaining both their geometric size and topological structure. However, during the propagation, the phase rotation of the vortex lattice induces a gradual rotation of the helicity in each skyrmionium unit cell. Notably, in contrast to prior plasmonic topological lattices of N\'eel-type skyrmions~\cite{tsesses2018optical,lei2021photonic}, our freespace topological optical lattices offer additional degrees of freedom for controlling the skyrmion helicity. This allows the realisation of Bloch-type skyrmions and antiskyrmion lattices, as shown in {\color{blue}Extended Data Fig.~2}. Furthermore, by superimposing two volumes of such topological optical lattice fields, we generate moiré-structured light fields that host even more complex topologies, as shown in Fig.~\ref{F1}(b). By varying the twist angle, the lattice topology, and the underlying symmetry, a broad family of nontrivial moiré patterns can be realised.

{

Depending on the twist angle, the resulting moiré field can form either commensurate (periodic) or incommensurate (aperiodic) geometries. The incommensurate patterns lack translational symmetry, however, the rotational symmetry is preserved. Commensurate patterns are formed when lattice points from the both constituent topological lattices coincide, resulting in well-defined periodicity. The commensurability condition can be expressed as~\cite{schwab2025skyrmion} 
\begin{equation}
  p\,\mathbf{a}_1 + q\,\mathbf{a}_2 \rightarrow q\,\mathbf{a}_1 + p\,\mathbf{a}_2,  
\end{equation}
where $p$ and $q$ are integers that define the coincidence between the primitive lattice vectors $\mathbf{a}_1$ and $\mathbf{a}_2$ of the two individual topological lattices. 

{
In the commensurate regime, distinct topological superlattices are formed at specific twist angles. For instance, at $\theta = 13.2^{\circ}$ the transverse field forms an optical superlattice composed of repeated supercells of complex skyrmionium bags — a topology not previously reported in any physical system (Fig.~\ref{F1}(c)). Each skyrmionium bag consists of multiple elementary skyrmionium cluster (e.g. $n = 7$) surrounded by a larger bag skyrmionium of opposite polarity. The outer bag skyrmionium offers an additional layer of topological protection for the embedded baby skyrmioniums. The central skyrmionium bag remains stable over a finite range of twist angles around $\theta = 13.2^{\circ}$, beyond which the long-range periodic order gradually disappears, while a single central skyrmionium bag is preserved {\color{blue}(see Supplementary Material)}.

{

Controlled variation in the twist angle leads to a variety of new supertopologies, including optical nested superlattices. Figure~\ref{F1}(d) shows a twisted bi-volume C$_6$ skyrmionium lattice field at $\theta = 27.8^{\circ}$, where a nondiffracting topological nested superlattice is formed in free space, where the skyrmioniums at the periphery are shared with neighboring super unit cells.  Our freespace optical nested superlattice show a new class of higher-order mixed topology, where the centre hosts a target skyrmion composed of multiple radially nested skyrmions ($n > 2$) with a total skyrmion number $|N_\text{sk}| = [1 + (-1)^n]/2$, surrounded by a set of skyrmioniums. Furthermore, for specific twist angle, freespace twistronics leads to quasicrystaline geometry similar to those recently reported in plasmonic systems~\cite{tsesses2025four,putley2025mixing,lin2025photonic}. Quasicrystals are ordered but non-periodic systems that exhibit rotational symmetry while lacking translational symmetry, yet can host nontrivial topological textures such as the skyrmion–meron mixed states observed in plasmonic quasicrystals~\cite{putley2025mixing}. For $\theta = 30^{\circ}$ freespace twistronics show a new class of higher-order mixed topology and can be termed as \textit{supertopological quasicrystals}  {\color{blue}(see Supplementary Material)}.
 
Furthermore, the individual C$_6$ skyrmionium lattices can be replaced with other topological lattices to obtain more diverse topologies. For instance, when C$_4$ skyrmion lattices are twisted, the moiré field at $\theta = 16.26^{\circ}$ exhibits a new type of topology not previously reported — the skyrmion cluster superlattice (Fig.~\ref{F1}(e)). The inset shows a skyrmion-cluster supercell that hosts $n$ skyrmions of identical topology with a total skyrmion number of $|N_\text{sk}| = n$. Similarly, for C$_6$ skyrmion lattices, the moiré field at $\theta = 13.2^{\circ}$ forms skyrmion bag superlattices (Fig.~\ref{F1}(f)). In a skyrmion bag, multiple elementary skyrmions ($n = 7$) are enclosed by a larger skyrmion bag of opposite polarity, giving a total skyrmion number of $|N_\text{sk}| = n - 1$ (see inset). Notably, the skyrmion bag topology was previously proposed in liquid crystals and magnetic systems~\cite{foster2019two}, and more recently observed in surface plasmon polaritons on metal surfaces~\cite{schwab2025skyrmion}. In contrast, the skyrmion bags demonstrated here are realised in freespace light fields that can propagate over arbitrarily long distances while preserving their topology.

\begin{figure*}
	\includegraphics[width=1\linewidth]{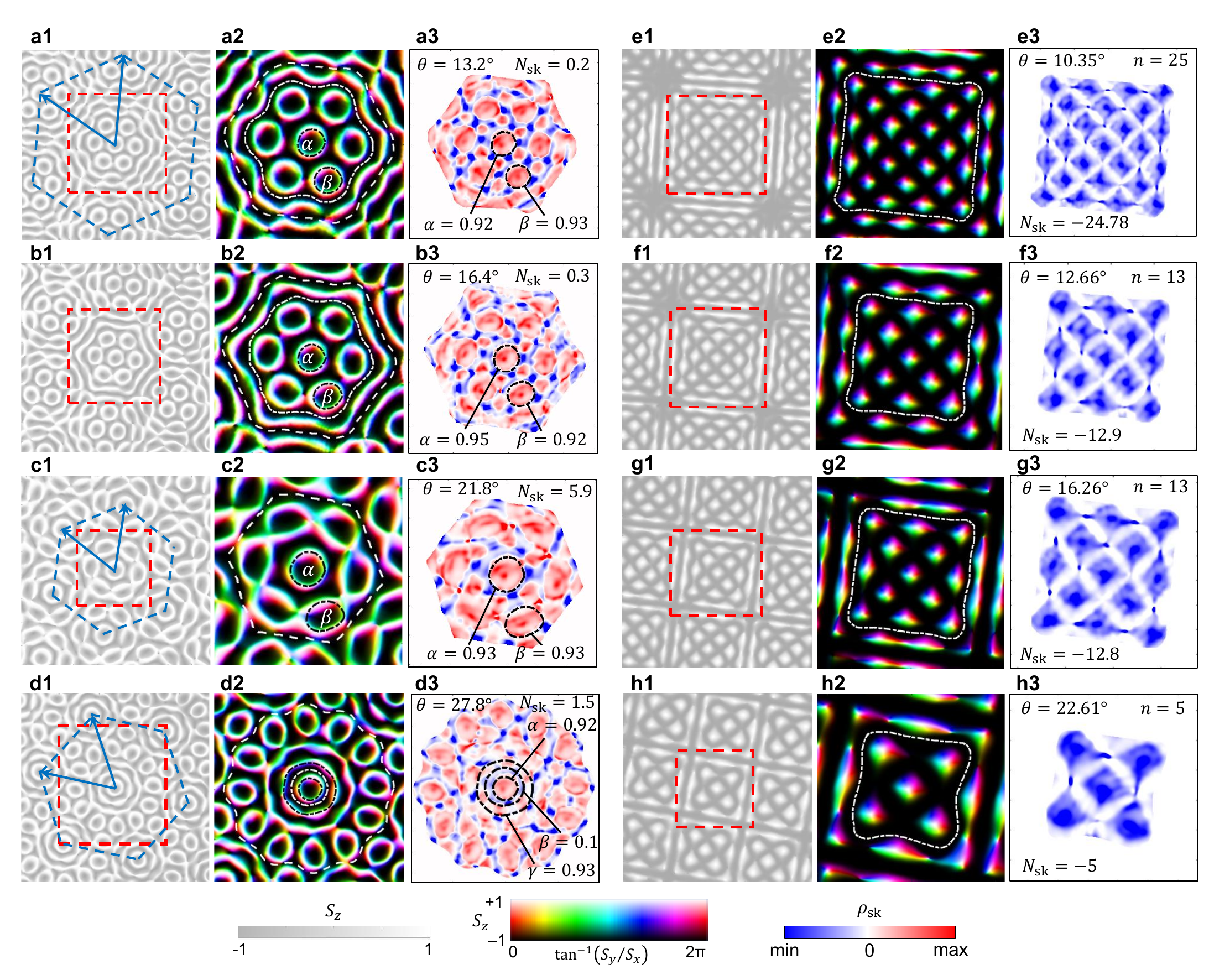}
	\caption{\textbf{Experimental results of optical supertopologies.} Long-range distributions of $S_z$, spin textures of featured topological structures (in red-dashed line boxes of long-range patterns correspondingly), and corresponding skyrmion density distributions for twisted C$_6$ skyrmionium lattices obtaining (a1-a3) skyrmionium bag superlattices at $\theta=13.2^{\circ}$ (blue arrows mark the long-range periodic unit vectors of supercell), (b1-b3) isolated skyrmionium bag at $\theta=16.4^{\circ}$, (c1-c3) skyrmion cluster superlattices at $\theta=21.8^{\circ}$, (d1-d3) nested superlattice at $\theta=27.8^{\circ}$; and for twisted C$_4$ skyrmion lattices obtaining skyrmion cluster superlattices for different baby skyrmion numbers in supercell of (e1-e3) $n=25$ at $\theta=10.35^{\circ}$, (f1-f3) $n=13$ at $\theta=12.66^{\circ}$, (g1-g3) $n=13$ at $\theta=16.26^{\circ}$, (h1-h3) $n=5$ at $\theta=22.61^{\circ}$. The experimental skyrmion numbers $N_\text{sk}$ of featured supertopologies, marked by big dashed lines in (a2-h2), are shown in (a3-h3), respectively. The experimental $N_\text{sk}$ values of selected protected baby skyrmions (``$\alpha$'',``$\beta$'',``$\gamma$'',) marked by black dashed lines are shown in (a3-d3), correspondingly.}
	\label{F2}
\end{figure*}

\begin{figure*}
	\includegraphics[width=0.9\linewidth]{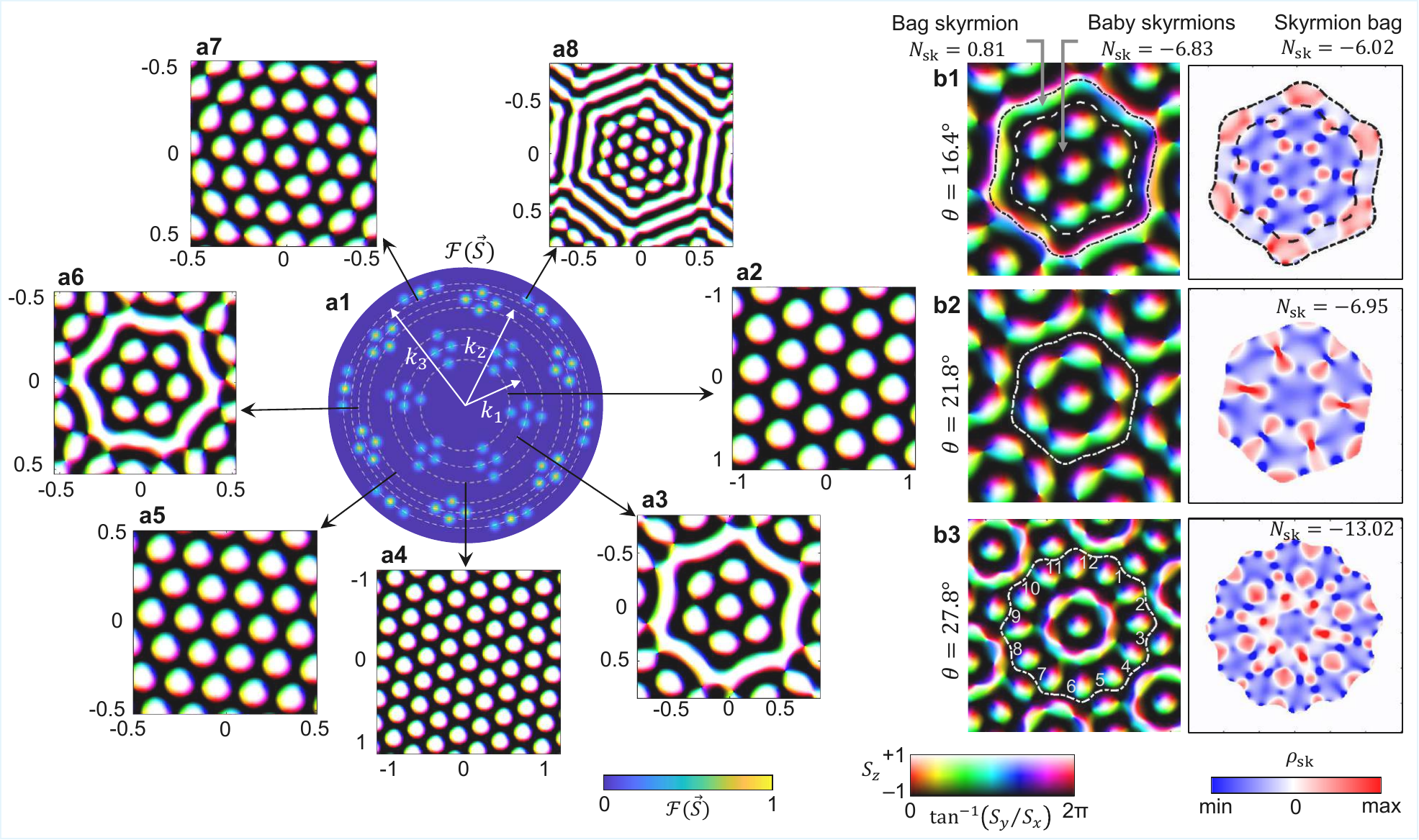}
	\caption{\textbf{Hierarchical supertopology decomposition.} (a1) Theoretical spatial Fourier spectrum of stokes vector of a skyrmionium bag superlattice at $\theta=16.4^{\circ}$, where the distribution is located at a set of concentric circles in wavenumber-space, and (a2-a8) the corresponding hierarchical real-space textures as skyrmion lattices and skyrmion bag superlattices. The white arrows in (a1) mark the three circles of wavenumber radii $k_1$, $k_2$, and $k_3$ for the cases of skyrmion bag generation. (b1-b3) Experimental results of skyrmion bag, skyrmion cluster superlattices, and topological quasicrystal, with measured textures and skyrmion density distributions, at twisted angles of $16.4^{\circ}$, $21.8^{\circ}$, and $27.8^{\circ}$, respectively, measured skyrmion numbers of selected topological featured in dashed lines are marked. Unit of coordinates: mm.}
	\label{F3}
\end{figure*}

\begin{figure*}[htbp]
	\includegraphics[width=0.95\linewidth]{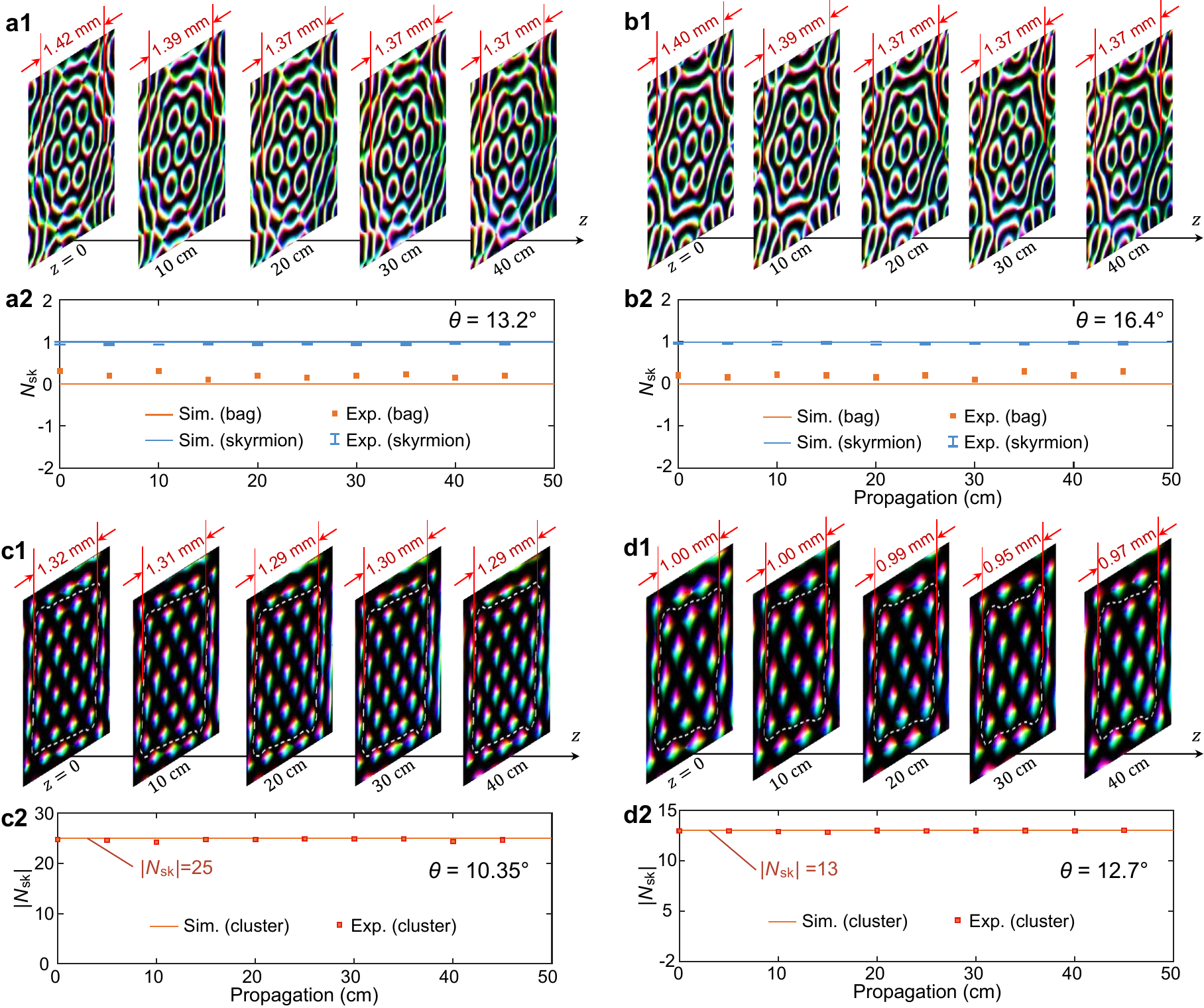}
	\caption{\textbf{Nondiffracting topologically stable propagation.} (a1-d1) Experimental results of the propagation-dependent supertopological textures for two skyrmionium bags at $\theta=13.2^{\circ}$ and $16.4^{\circ}$ and two skyrmion clusters at $\theta=10.35^{\circ}$ and $12.7^{\circ}$, respectively, form $z=0$ to 50 cm. The featured side is marked correspondingly in each texture pattern. (a2,b2) The simulated and experimental skyrmion numbers for the skyrmionium bags ($n=7$) and the protected baby skyrmions in which (the error bars evaluate the numerical difference for the 7 baby skyrmions), corresponding to (a1,b1). (c2,d2) The simulated and experimental skyrmion numbers for the skyrmion clusters ($n=25$ and $n=13$), corresponding to (c1,d1).}
	\label{F4}
\end{figure*}

\begin{figure*}
	\includegraphics[width=0.9\linewidth]{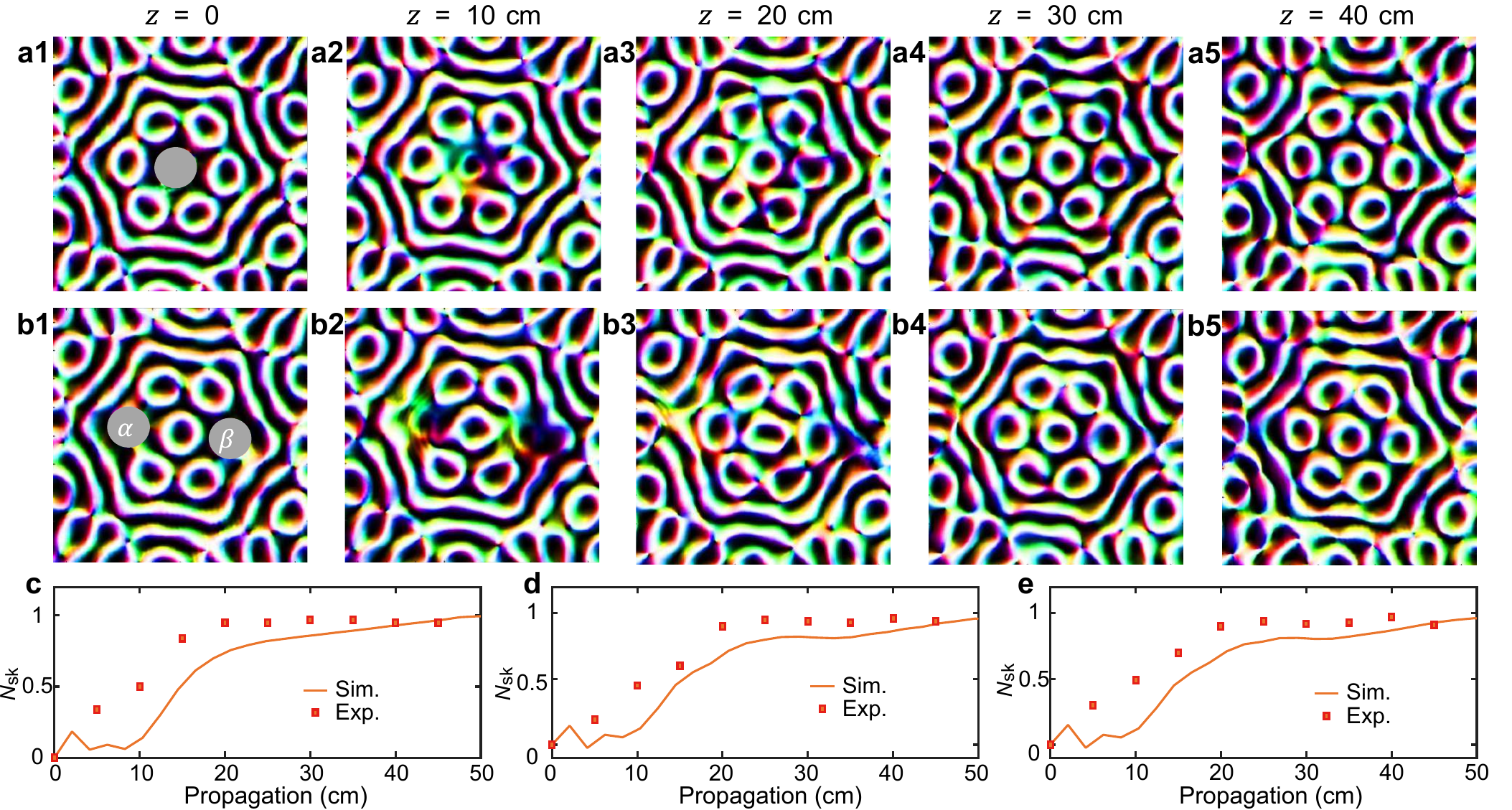}
	\caption{\textbf{Self-healing topologies against obstacles.} Experimental results of skyrmionium bag propagation passing through obstacles for (a1) a disk, marked in gray, blocking the center skyrmionium and (a2) two disks blocking two skyrmioniums in bag, at $z=0$. (c-e) The simulated and experimental skyrmion number versus propagation for the baby skyrmion protected in the blocked skyrmionium in (a1) and that two in the two blocked skyrmioniums (``$\alpha$'' and ``$\beta$'') in (b1), respectively.}
	\label{F5}
\end{figure*}

\section{Results of Supertopologies}
To show the proof of concept, we experimentally realise the freespace supertopologies using a spatial light modulator (SLM) {\color{blue}(see  Supplementary Material)}. We first present the results for twisted C$_6$-based skyrmionium lattices at $\theta = 13.2^{\circ}$, where superlattices of skyrmionium bags act as supercells arranged in long-range C$_6$ symmetry. Figure~\ref{F2}(a1) shows the measured long-range order in the out-of-plane Stokes vector $S_z$, where the arrows indicate the unit vectors defining the hexagonal supercells formed by the skyrmionium bags. The polarisation texture of the central skyrmionium bag is shown in Fig.~\ref{F2}(a2), where the white dash–dot line marks the boundary of the skyrmionium cluster formed by seven individual skyrmioniums. This cluster is enclosed within a larger skyrmionium bag of opposite polarity, whose boundary is outlined by the dashed white curve. The boundaries are identified using a computational Snake algorithm {\color{blue}(see Methods)}. Figure~\ref{F2}(a3) shows the measured skyrmion density for the central skyrmionium bag. As evident, the elementary skyrmioniums within the cluster share the same vorticity as the surrounding bag skyrmionium but with opposite polarity. The measured skyrmion numbers of the entire skyrmionium bag and its inner skyrmion cluster are both close to zero, whereas each baby skyrmion embedded in the cluster exhibits a near-unit topology ($N_\text{sk} > 0.9$).

As the twist angle increases to $\theta = 16.4^{\circ}$, the overall long range C$_6$ symmetry begins to break down. Figure~\ref{F2}(b1) shows that while the central skyrmionium bag remains intact, the surrounding region becomes aperiodic and no longer forms a regular superlattice. The corresponding polarisation and skyrmion density maps (Figs.~\ref{F2}(b2,b3)) confirm that the core topology of the central skyrmionium bag is preserved despite the loss of long-range order, demonstrating its resilience against twist angle deviation. Further increasing the twist angle to $\theta = 21.8^{\circ}$ leads to a topological transition into the skyrmion cluster superlattice, as shown in Fig.~\ref{F2}(c1). In this configuration, the supercells adopt a petal-like arrangement consisting of multiple skyrmion clusters surrounding a central skyrmionium. The polarisation and skyrmion density maps (Figs.~\ref{F2}(c2,c3)) show that the central skyrmionium and its protected skyrmion exhibit higher structural fidelity compared to the six surrounding petals.

Furthermore, at $\theta=27.8^{\circ}$, a nested superlattices emerges, where the basic central structure is a multi-$\pi$ nested target skyrmion surrounded with multiple skyrmionium tiling the whole space, with global transitional symmetry (Fig.~\ref{F2}(d1)). The first three nested skyrmion at the center with measured skyrmion numbers close to $\pm1$ were analyzed in Figs.~\ref{F2}(d2,d3)


Furthermore, we generate diverse moiré supertopologies by performing freespace twistronics on C$_4$-based skyrmion lattices, resulting in skyrmion cluster superlattices that contain different numbers of skyrmions per supercell at different twist angles. At $\theta = 10.35^{\circ}$, the moiré superlattice supercell contains 25 skyrmions (Figs.~\ref{F2}(e1–e3)). When the twist angle is increased to $\theta = 12.66^{\circ}$, the structure transforms into a 13-skyrmion supercell (Figs.~\ref{F2}(f1–f3)), which remains stable up to $\theta = 16.26^{\circ}$ (Figs.~\ref{F2}(g1–g3)). Further increase to $\theta = 22.61^{\circ}$ results in a 5-skyrmion supercell (Figs.~\ref{F2}(h1–h3)). For the C$_4$-based skyrmion lattices, there is no one-to-one mapping between the vortices and Gaussians in the left- and right-circularly polarised (LCP and RCP) components, leading to intensity nulls or electromagnetic dark spots within the lattice. When such lattices are twisted, these intensity nulls persist in the resulting moiré superlattices, which can introduce small inaccuracies in the measurement of total skyrmion number. To ensure accurate measurements, a small correction term $\delta$ is added to the LCP component. Furthermore, freespace twistronics is also performed on C$_3$-based meron lattices, yielding complex meron–antimeron cluster supertopologies {\color{blue}(see Supplementary Material)}.

Moreover, it has been shown that due to periodicity in the stokes vector, the topological lattices can be decomposed in sub-lattices of lower topologies. For example, a C$_6$ skyrmionium lattice can be decomposed in three different C$_6$ skyrmion sub-lattices of different periodicity. Similarly, a C$_4$ skyrmion lattice can be decomposed in two C$_4$ meron sub-lattices ~\cite{teng2025topological}. Therefore, in addition to the supertopologies discussed above, we can also achieve skyrmion bags, skyrmion petal superlattices, lower-order topological quasicrystals, and so on. To observe freespace skyrmion bag superlattices, we need the optical lattice decomposition method~\cite{teng2025topological}. Similar to the Stokes vector of individual C$_6$ skyrmionium lattices, the Stokes vector of twisted superlattice also exhibits periodicity, leading to various spatial frequency components. Therefore, by selective spatial decomposition, lower order topologies can be obtained. Figure~\ref{F3}(a1) shows the spatial Fourier spectrum of the Stokes vector (Fig.~\ref{F2}(b1)) corresponding to} skyrmionium bag superlattice obtained by twisting two elementary skyrmionium lattices at an angle of  $\theta=16.4^{\circ}$. The Fourier spectrum can be divided into seven sets of spatial frequency components (shown by white dashed circles in Fig.~\ref{F3}(a1)). Figures~\ref{F3}(a2-a8) show 7 real-space textures which are obtained from the seven different spatial frequency circles. These textures resemble with C$_6$ skyrmion lattices, as well as skyrmion bags in bilayer and trilayer moiré skyrmion superlattices. Spatial frequencies consisting of only six spots results in C$_6$ skyrmion lattices in real space with different lattice vectors (Figs.~\ref{F3}(a2-a8). However, sets of spatial frequencies consisting of more than six spots, i.e. $k_1$ and $k_2$ with twelve and $k_3$ with eighteen spots results in higher order super topologies. Note, $k_1$ and $k_2$ results in skyrmion bag corresponding consisting of seven elementary skyrmions (Fig.~\ref{F3}(a3)-(a6)), whereas, $k_3$ generates a bigger skyrmion bag with eighteen elementary skyrmions (Fig.~\ref{F3}(a8)). Figure~\ref{F3}(b1) shows the experimental results of the skyrmion bag, where the $n=7$ baby skyrmions are inside with measured $N_\text{sk}=-6.83$ surrounded with a bag skyrmion of opposite polarity with $N_\text{sk}=0.81$, thus, the total skyrmion number is $N_\text{sk}=-6.83+0.81=-6.02$ and its absolute value is close to the idea $n-1=6$ (Fig.~\ref{F3}(b1)). If we do same decomposition to mori\'e textures at $\theta=21.8^{\circ}$, we can obtain skyrmion cluster superlattice, where the supercell is petal like cluster of 7 skyrmions (Fig.~\ref{F3}(b2)). At $\theta=27.8^{\circ}$, the decomposed mori\'e texture shows a lower-order topological nested superlattices, where the basic structure is a $3\pi$ target skyrmion surrounded with 12 skyrmions, thus the total skyrmion number is close to $|N_\text{sk}|=13$ (Fig.~\ref{F3}(b3)). We can apply such decomposition to other mori\'e supertopologies to access more kinds of topologies.

\section{Robustness upon propagation}
The supertopologies are generated by twisting the light fields of two elementary topological lattices. Therefore, they exhibit the propagation properties of elementary lattices. For example, light field of elementary C$_6$ skyrmionium lattice is composed of only one radial frequency component with six point sources. Therefore, the light field from six point sources diffracts equally and results in non-diffracting and propagation invariant behavior in the interference region. In contrast to conventional skyrmionic or bimeronic beams where they are divergent but topologically invariant upon propagation~\cite{shen2021topological}, here, the topological features and geometric shape and size the supertopologies all do not change at all. 
Figures~\ref{F4}(a,b) show experimentally measured skyrmion bag textures upon propagation through 50 cm, in twisted C$_6$ skyrmionium lattices at $\theta=13.2^{\circ}$ (superlattice case) and $16.4^{\circ}$ (isolated case). In both cases, the geometric sizes and topological features are invariant, only with local helicity evolves with dynamic phase upon propagation, \YS{see details in Extended data figure E7-E8}.
Figures~\ref{F4}(a2,b2) are corresponding simulated and experimental skyrmion numbers for the skyrmionium bags ($n=7$) and the protected baby skyrmions in which (the error bars evaluate the numerical difference for the 7 baby skyrmions). The results prove that the supertopology is resilient to both dynamic propagation and twisted angle deviations.

Figures~\ref{F4}(c,d) show experimentally measured nondiffractive propagation-dependent geometry-independent skyrmion clusters, in twisted C$_4$ skyrmion lattices at $\theta=10.35^{\circ}$ and $12.7^{\circ}$ for $n=25$ and $n=13$, respectively. The measured skyrmion numbers agree very well with the theoretical prediction, see Figs~\ref{F4}(c1,d1). \YS{Please see details in Extended data figure E7-E8.}

Hereinafter, we show the nondiffractive supertopologies are even strongly resilient to obstacles. Notably, the topological protection of skyrmions were previously studied~\cite{wang2024topological,zhang2025topological}, however, which are only for complex transparent and scattering media, and no similar studies to topological lattices is proposed. The blocking of solid obstacles are of stronger perturbation in contrast to complex media, the optical mode revival after obstacles in propagation path is usually called self-healing effect~\cite{shen2022self}, the topology self-healing of skyrmionic beam was also studied recently~\cite{guo2025self}, but again, not for topological lattice.

Figures~\ref{F5}(a,b) show our experimental results of skyrmionium bag propagation passing through obstacles of consisting of one disk (a) and two disks (b), blocking one and two skyrmioniums in the bag, respectively. The simulated and experimental skyrmion number versus propagation for the baby skyrmion protected in the blocked skyrmionium and that two in the two blocked skyrmioniums (``$\alpha$'' and ``$\beta$'') are shown in Fig.~\ref{F5}(c-e), respectively.
\YS{Please see details in Extended data figure E9.}

\section{Discussion}

We proposed the concept of freespace twistronics and successfully demonstrated emerging supertopologies controlled in the twisted mori\'e structured light, opening many exciting possibilities of new directions.
Many novel topologies, skyrmionium bags, skyrmion-bag superlattices, supertopological quasicrystals, etc., not accessible in any physical system before, now, can be simply generated in twisted bi-volume mori\'e structured light. In this work, we only considered superposition of two layers. It can be extended to mori\'e-of-mori\'e scheme as recently emerging trilayer lattices~\cite{park2021tunable,park2025unconventional} and twisted multilayer~\cite{torma2022superconductivity} for opening unlimited exploration of more nontrivial topologies.

Breaking 2D limit of prior twistronics, we open new dimensions and more degrees of freedom for 3D even 4D space-time twistronics. In this work, we only considered coaxial twisting, while, for emerging 3D freespace topological optical field, e.g. the space-time hopfion crystals~\cite{lin2025space}, we can open 3D twisting of multiple topological along arbitrary axial orientations for potential higher-dimensional space-time supercrystals.

Here, we focus on exploring topological textures of light. More properties of structured light can be studied in the framework of freespace twistronics, such as energy localization~\cite{wang2020localization} and superoscillation~\cite{ma2024observation}.

In contrast to any prior model of mori\'e electronics and photonics relying on responses of materials, our freespace mori\'e supertopological light can propagate so as to interact with matter, potentially transfer optical topologies to matter, as a new methodology to excite physical effects, atomic topologies~\cite{mitra2025topological}, for atom trapping~\cite{yu2024observing,meng2023atomic}, structured light directly writing topological solitons of matter.

Optical supertopologies are highly ondemand for next-generation information carriers to revolutionize our optical communication network~\cite{wan2023ultra}. As their diversified control of topological textures, higher capacity data encoding and transfer can be promised, meanwhile, the enhanced robustness can be expected due to the nondiffraction and resilience of supertopologies. We plan to apply supertopologies for communication and encryption, overcoming the bottleneck of current optical vortex techniques.

Supertopologies were firstly generated in light waves here, which will certainly motivate the exploration of supertopologies in other waves, e.g. sound waves and water waves, as well as the exploration of supertopologies in matter, i.e. liquid crystals, chiral magnets and beyond.


\vspace{0.8em}\noindent  \textbf{Data availability}

\noindent{The data that support the findings of this study are available within the paper and the Extended Data. Other relevant data are available from the corresponding authors on reasonable request. 

\vspace{0.8em}\noindent  \textbf{Acknowledgements}

\noindent{Authors thank Julian Schwab for useful discussions. Authors thank Singapore Ministry of Education (MOE) AcRF Tier 1 grants (RG157/23 \& RT11/23), Singapore Agency for Science, Technology and Research (A*STAR) MTC Individual Research Grants (M24N7c0080), and Nanyang Assistant Professorship Start Up grant.}

\vspace{0.8em}\noindent  \textbf{Author Contributions}

\noindent Y.~S. conceived the idea and supervised the project; V.~D. developed the theoretical simulation and experiments; All authors analyzed the data and prepared the manuscript.

\vspace{0.8em}\noindent  \textbf{Conflict of Interest}

\noindent The authors declare no conflict of interest.

\bibliographystyle{naturemag}
\bibliography{bibfile-new}

@article{carr2017twistronics,
  title={Twistronics: Manipulating the electronic properties of two-dimensional layered structures through their twist angle},
  author={Carr, Stephen and Massatt, Daniel and Fang, Shiang and Cazeaux, Paul and Luskin, Mitchell and Kaxiras, Efthimios},
  journal={Physical Review B},
  volume={95},
  number={7},
  pages={075420},
  year={2017},
  publisher={APS}
}

@article{wu2025optical,
  title={Optical skyrmion lattices accelerating in a free-space mode},
  author={Wu, Haijun and Zhou, Weijie and Zhu, Zhihan and Shen, Yijie},
  journal={APL Photonics},
  volume={10},
  number={5},
  year={2025},
  publisher={AIP Publishing}
}

@article{shen2021topological,
  title={Topological bimeronic beams},
  author={Shen, Yijie},
  journal={Optics Letters},
  volume={46},
  number={15},
  pages={3737--3740},
  year={2021},
  publisher={OSA}
}

@article{ma2024observation,
  title={Observation of Superoscillation Superlattices},
  author={Ma, Xin and Zhang, Hao and Wei, Wenjun and Tai, Yuping and Li, Xinzhong and Shen, Yijie},
  journal={arXiv preprint arXiv:2409.19565},
  year={2024}
}

@article{kass1988snakes,
  title={Snakes: Active contour models},
  author={Kass, Michael and Witkin, Andrew and Terzopoulos, Demetri},
  journal={International journal of computer vision},
  volume={1},
  number={4},
  pages={321--331},
  year={1988},
  publisher={Springer}
}

@article{ouyang2024singular,
  title={Singular dielectric nanolaser with atomic-scale field localization},
  author={Ouyang, Yun-Hao and Luan, Hong-Yi and Zhao, Zi-Wei and Mao, Wen-Zhi and Ma, Ren-Min},
  journal={Nature},
  volume={632},
  number={8024},
  pages={287--293},
  year={2024},
  publisher={Nature Publishing Group UK London}
}

@article{hu2020topological,
  title={Topological polaritons and photonic magic angles in twisted $\alpha$-MoO3 bilayers},
  author={Hu, Guangwei and Ou, Qingdong and Si, Guangyuan and Wu, Yingjie and Wu, Jing and Dai, Zhigao and Krasnok, Alex and Mazor, Yarden and Zhang, Qing and Bao, Qiaoliang and others},
  journal={Nature},
  volume={582},
  number={7811},
  pages={209--213},
  year={2020},
  publisher={Nature Publishing Group UK London}
}

@article{lin2025space,
  title={Space-Time Optical Hopfion Crystals},
  author={Lin, Wenbo and Mata-Cervera, Nilo and Ota, Yasutomo and Shen, Yijie and Iwamoto, Satoshi},
  journal={Physical Review Letters},
  volume={135},
  number={8},
  pages={083801},
  year={2025},
  publisher={APS}
}

@article{wang2024topological,
  title={Topological protection of optical skyrmions through complex media},
  author={Wang, An Aloysius and Zhao, Zimo and Ma, Yifei and Cai, Yuxi and Zhang, Runchen and Shang, Xiaoyi and Zhang, Yunqi and Qin, Ji and Pong, Zhi-Kai and Marozs{\'a}k, T{\'a}d{\'e} and others},
  journal={Light: Science \& Applications},
  volume={13},
  number={1},
  pages={314},
  year={2024},
  publisher={Nature Publishing Group UK London}
}

@article{zhang2025topological,
  title={Topological protection degrees of optical skyrmions and their electrical control},
  author={Zhang, Zan and Xie, Xi and Zhuang, Chuhong and Wu, Binyu and Liu, Zihan and Wu, Baoyun and Mihalache, Dumitru and Shen, Yijie and Deng, Dongmei and others},
  journal={Photonics Research},
  volume={13},
  number={9},
  pages={B1--B11},
  year={2025},
  publisher={Editorial Office of Photonics Research}
}

@article{lei2021photonic,
  title={Photonic spin lattices: symmetry constraints for skyrmion and meron topologies},
  author={Lei, Xinrui and Yang, Aiping and Shi, Peng and Xie, Zhenwei and Du, Luping and Zayats, Anatoly V and Yuan, Xiaocong},
  journal={Physical Review Letters},
  volume={127},
  number={23},
  pages={237403},
  year={2021},
  publisher={APS}
}

@article{guo2025self,
  title={Self-healing of optical skyrmionic beams},
  author={Guo, Haochen and Das, Trishita and Wu, Haijun and Dev, Vasu and Zhu, Zhihan and Shen, Yijie},
  journal={Journal of Optics},
  volume={27},
  number={2},
  pages={025604},
  year={2025},
  publisher={IOP Publishing}
}

@article{shen2022self,
  title={Self-healing of structured light: a review},
  author={Shen, Yijie and Pidishety, Shankar and Nape, Isaac and Dudley, Angela},
  journal={Journal of Optics},
  volume={24},
  number={10},
  pages={103001},
  year={2022},
  publisher={IOP Publishing}
}

@article{foster2019two,
  title={Two-dimensional skyrmion bags in liquid crystals and ferromagnets},
  author={Foster, David and Kind, Charles and Ackerman, Paul J and Tai, Jung-Shen B and Dennis, Mark R and Smalyukh, Ivan I},
  journal={Nature Physics},
  volume={15},
  number={7},
  pages={655--659},
  year={2019},
  publisher={Nature Publishing Group UK London}
}

@article{teng2025topological,
  title={Topological decomposition of hierarchical skyrmion lattices},
  author={Teng, Houan and Zhong, Jinzhan and Lei, Xinrui and Zhan, Qiwen},
  journal={Communications Physics},
  volume={8},
  number={1},
  pages={99},
  year={2025},
  publisher={Nature Publishing Group UK London}
}

@article{lin2025photonic,
  title={Photonic quasicrystal of spin angular momentum},
  author={Lin, Min and Gou, Xinxin and Xie, Zhenwei and Yang, Aiping and Du, Luping and Yuan, Xiaocong},
  journal={Science Advances},
  volume={11},
  number={18},
  pages={eadv3938},
  year={2025},
  publisher={American Association for the Advancement of Science}
}

@article{putley2025mixing,
  title={Mixing skyrmions and merons in topological quasicrystals of the evanescent optical field},
  author={Putley, Henry J and Davies, Bryn and Rodr{\'\i}guez-Fortu{\~n}o, Francisco J and Bykov, A Yu and Zayats, Anatoly V},
  journal={Optica},
  volume={12},
  number={5},
  pages={614--619},
  year={2025},
  publisher={Optica Publishing Group}
}

@article{tsesses2025four,
  title={Four-dimensional conserved topological charge vectors in plasmonic quasicrystals},
  author={Tsesses, Shai and Dreher, Pascal and Janoschka, David and Neuhaus, Alexander and Cohen, Kobi and Meiler, Tim C and Bucher, Tomer and Sapir, Shay and Frank, Bettina and Davis, Timothy J and others},
  journal={Science},
  volume={387},
  number={6734},
  pages={644--648},
  year={2025},
  publisher={American Association for the Advancement of Science}
}

@article{bogdanov2020physical,
  title={Physical foundations and basic properties of magnetic skyrmions},
  author={Bogdanov, Alexei N and Panagopoulos, Christos},
  journal={Nature Reviews Physics},
  volume={2},
  number={9},
  pages={492--498},
  year={2020},
  publisher={Nature Publishing Group UK London}
}

@article{shen2024optical,
  title={Optical skyrmions and other topological quasiparticles of light},
  author={Shen, Yijie and Zhang, Qiang and Shi, Peng and Du, Luping and Yuan, Xiaocong and Zayats, Anatoly V},
  journal={Nature Photonics},
  volume={18},
  number={1},
  pages={15--25},
  year={2024},
  publisher={Nature Publishing Group UK London}
}

@article{marco2024propagation,
  title={Propagation-invariant optical meron lattices},
  author={Marco, David and Herrera, Isael and Brasselet, Sophie and Alonso, Miguel A},
  journal={ACS Photonics},
  volume={11},
  number={6},
  pages={2397--2405},
  year={2024},
  publisher={ACS Publications}
}

@article{marco2024periodic,
  title={Periodic skyrmionic textures via conformal cartographic projections},
  author={Marco, David and Herrera, Isael and Brasselet, Sophie and Alonso, Miguel A},
  journal={APL Photonics},
  volume={9},
  number={11},
  year={2024},
  publisher={AIP Publishing}
}

@article{tsesses2018optical,
  title={Optical skyrmion lattice in evanescent electromagnetic fields},
  author={Tsesses, Shai and Ostrovsky, Evgeny and Cohen, Kobi and Gjonaj, Bergin and Lindner, Netanel H and Bartal, Guy},
  journal={Science},
  volume={361},
  number={6406},
  pages={993--996},
  year={2018},
  publisher={American Association for the Advancement of Science}
}

@article{mitra2025topological,
  title={Topological optical skyrmion transfer to matter},
  author={Mitra, Chirantan and Madasu, Chetan Sriram and Gabardos, Lucas and Kwong, Chang Chi and Shen, Yijie and Ruostekoski, Janne and Wilkowski, David},
  journal={APL Photonics},
  volume={10},
  number={4},
  year={2025},
  publisher={AIP Publishing}
}

@article{yu2024observing,
  title={Observing the two-dimensional Bose glass in an optical quasicrystal},
  author={Yu, Jr-Chiun and Bhave, Shaurya and Reeve, Lee and Song, Bo and Schneider, Ulrich},
  journal={Nature},
  volume={633},
  number={8029},
  pages={338--343},
  year={2024},
  publisher={Nature Publishing Group UK London}
}

@article{wu2025photonic,
  title={Photonic Torons with 3D Topology Transitions and Tunable Spin Monopoles},
  author={Wu, Haijun and Mata-Cervera, Nilo and Wang, Haiwen and Zhu, Zhihan and Qiu, Chengwei and Shen, Yijie},
  journal={Physical Review Letters},
  volume={135},
  number={6},
  pages={063802},
  year={2025},
  publisher={APS}
}

@article{wu2022hopfions,
  title={Hopfions, heliknotons, skyrmions, torons and both abelian and nonabelian vortices in chiral liquid crystals},
  author={Wu, Jin-Sheng and Smalyukh, Ivan I},
  journal={Liquid Crystals Reviews},
  volume={10},
  number={1-2},
  pages={34--68},
  year={2022},
  publisher={Taylor \& Francis}
}

@article{zheng2023hopfion,
  title={Hopfion rings in a cubic chiral magnet},
  author={Zheng, Fengshan and Kiselev, Nikolai S and Rybakov, Filipp N and Yang, Luyan and Shi, Wen and Bl{\"u}gel, Stefan and Dunin-Borkowski, Rafal E},
  journal={Nature},
  volume={623},
  number={7988},
  pages={718--723},
  year={2023},
  publisher={Nature Publishing Group UK London}
}

@article{tang2021magnetic,
  title={Magnetic skyrmion bundles and their current-driven dynamics},
  author={Tang, Jin and Wu, Yaodong and Wang, Weiwei and Kong, Lingyao and Lv, Boyao and Wei, Wensen and Zang, Jiadong and Tian, Mingliang and Du, Haifeng},
  journal={Nature Nanotechnology},
  volume={16},
  number={10},
  pages={1086--1091},
  year={2021},
  publisher={Nature Publishing Group UK London}
}

@article{jani2021antiferromagnetic,
  title={Antiferromagnetic half-skyrmions and bimerons at room temperature},
  author={Jani, Hariom and Lin, Jheng-Cyuan and Chen, Jiahao and Harrison, Jack and Maccherozzi, Francesco and Schad, Jonathon and Prakash, Saurav and Eom, Chang-Beom and Ariando, Ariando and Venkatesan, Thirumalai and others},
  journal={Nature},
  volume={590},
  number={7844},
  pages={74--79},
  year={2021},
  publisher={Nature Publishing Group UK London}
}

@article{ciarrocchi2022excitonic,
  title={Excitonic devices with van der Waals heterostructures: valleytronics meets twistronics},
  author={Ciarrocchi, Alberto and Tagarelli, Fedele and Avsar, Ahmet and Kis, Andras},
  journal={Nature Reviews Materials},
  volume={7},
  number={6},
  pages={449--464},
  year={2022},
  publisher={Nature Publishing Group UK London}
}

@article{shen2025free,
  title={Free-space topological optical textures: tutorial},
  author={Shen, Yijie and Wang, Haiwen and Fan, Shanhui},
  journal={Advances in Optics and Photonics},
  volume={17},
  number={2},
  pages={295--374},
  year={2025},
  publisher={Optica Publishing Group}
}

@article{cao2020tunable,
  title={Tunable correlated states and spin-polarized phases in twisted bilayer--bilayer graphene},
  author={Cao, Yuan and Rodan-Legrain, Daniel and Rubies-Bigorda, Oriol and Park, Jeong Min and Watanabe, Kenji and Taniguchi, Takashi and Jarillo-Herrero, Pablo},
  journal={Nature},
  volume={583},
  number={7815},
  pages={215--220},
  year={2020},
  publisher={Nature Publishing Group UK London}
}

@article{cao2018correlated,
  title={Correlated insulator behaviour at half-filling in magic-angle graphene superlattices},
  author={Cao, Yuan and Fatemi, Valla and Demir, Ahmet and Fang, Shiang and Tomarken, Spencer L and Luo, Jason Y and Sanchez-Yamagishi, Javier D and Watanabe, Kenji and Taniguchi, Takashi and Kaxiras, Efthimios and others},
  journal={Nature},
  volume={556},
  number={7699},
  pages={80--84},
  year={2018},
  publisher={Nature Publishing Group UK London}
}

@article{wang2020localization,
  title={Localization and delocalization of light in photonic moir{\'e} lattices},
  author={Wang, Peng and Zheng, Yuanlin and Chen, Xianfeng and Huang, Changming and Kartashov, Yaroslav V and Torner, Lluis and Konotop, Vladimir V and Ye, Fangwei},
  journal={Nature},
  volume={577},
  number={7788},
  pages={42--46},
  year={2020},
  publisher={Nature Publishing Group UK London}
}

@article{cao2018unconventional,
  title={Unconventional superconductivity in magic-angle graphene superlattices},
  author={Cao, Yuan and Fatemi, Valla and Fang, Shiang and Watanabe, Kenji and Taniguchi, Takashi and Kaxiras, Efthimios and Jarillo-Herrero, Pablo},
  journal={Nature},
  volume={556},
  number={7699},
  pages={43--50},
  year={2018},
  publisher={Nature Publishing Group UK London}
}

@article{park2021tunable,
  title={Tunable strongly coupled superconductivity in magic-angle twisted trilayer graphene},
  author={Park, Jeong Min and Cao, Yuan and Watanabe, Kenji and Taniguchi, Takashi and Jarillo-Herrero, Pablo},
  journal={Nature},
  volume={590},
  number={7845},
  pages={249--255},
  year={2021},
  publisher={Nature Publishing Group UK London}
}

@article{meng2023atomic,
  title={Atomic Bose--Einstein condensate in twisted-bilayer optical lattices},
  author={Meng, Zengming and Wang, Liangwei and Han, Wei and Liu, Fangde and Wen, Kai and Gao, Chao and Wang, Pengjun and Chin, Cheng and Zhang, Jing},
  journal={Nature},
  volume={615},
  number={7951},
  pages={231--236},
  year={2023},
  publisher={Nature Publishing Group UK London}
}

@article{ji2024opto,
  title={Opto-twistronic Hall effect in a three-dimensional spiral lattice},
  author={Ji, Zhurun and Zhao, Yuzhou and Chen, Yicong and Zhu, Ziyan and Wang, Yuhui and Liu, Wenjing and Modi, Gaurav and Mele, Eugene J and Jin, Song and Agarwal, Ritesh},
  journal={Nature},
  volume={634},
  number={8032},
  pages={69--73},
  year={2024},
  publisher={Nature Publishing Group UK London}
}

@article{park2025unconventional,
  title={Unconventional domain tessellations in moir{\'e}-of-moir{\'e} lattices},
  author={Park, Daesung and Park, Changwon and Yananose, Kunihiro and Ko, Eunjung and Kim, Byunghyun and Engelke, Rebecca and Zhang, Xi and Davydov, Konstantin and Green, Matthew and Kim, Hyun-Mi and others},
  journal={Nature},
  pages={1--8},
  year={2025},
  publisher={Nature Publishing Group}
}

@article{schwab2025skyrmion,
  title={Skyrmion bags of light in plasmonic moir{\'e} superlattices},
  author={Schwab, Julian and Neuhaus, Alexander and Dreher, Pascal and Tsesses, Shai and Cohen, Kobi and Mangold, Florian and Mantha, Anant and Frank, Bettina and Bartal, Guy and Meyer zu Heringdorf, Frank-J and others},
  journal={Nature Physics},
  pages={1--7},
  year={2025},
  publisher={Nature Publishing Group UK London}
}

@article{luan2023reconfigurable,
  title={Reconfigurable moir{\'e} nanolaser arrays with phase synchronization},
  author={Luan, Hong-Yi and Ouyang, Yun-Hao and Zhao, Zi-Wei and Mao, Wen-Zhi and Ma, Ren-Min},
  journal={Nature},
  volume={624},
  number={7991},
  pages={282--288},
  year={2023},
  publisher={Nature Publishing Group UK London}
}

@article{du2023moire,
  title={Moir{\'e} photonics and optoelectronics},
  author={Du, Luojun and Molas, Maciej R and Huang, Zhiheng and Zhang, Guangyu and Wang, Feng and Sun, Zhipei},
  journal={Science},
  volume={379},
  number={6639},
  pages={eadg0014},
  year={2023},
  publisher={American Association for the Advancement of Science}
}

@article{xu2024hydrodynamic,
  title={Hydrodynamic moir{\'e} superlattice},
  author={Xu, Guoqiang and Zhou, Xue and Chen, Weijin and Hu, Guangwei and Yan, Zhiyuan and Li, Zhipeng and Yang, Shuihua and Qiu, Cheng-Wei},
  journal={Science},
  volume={386},
  number={6728},
  pages={1377--1383},
  year={2024},
  publisher={American Association for the Advancement of Science}
}

@article{torma2022superconductivity,
  title={Superconductivity, superfluidity and quantum geometry in twisted multilayer systems},
  author={T{\"o}rm{\"a}, P{\"a}ivi and Peotta, Sebastiano and Bernevig, Bogdan A},
  journal={Nature Reviews Physics},
  volume={4},
  number={8},
  pages={528--542},
  year={2022},
  publisher={Nature Publishing Group UK London}
}

@article{wan2023ultra,
  title={Ultra-degree-of-freedom structured light for ultracapacity information carriers},
  author={Wan, Zhensong and Wang, Hao and Liu, Qiang and Fu, Xing and Shen, Yijie},
  journal={ACS Photonics},
  volume={10},
  number={7},
  pages={2149--2164},
  year={2023},
  publisher={ACS Publications}
}

\section{methods}
\noindent \textbf{Characterization of skyrmion and its boundary}
The topological properties of the optical quantum skyrmions can be characterized by the skyrmion number, which is defined as\cite{shen2024optical}
\[
N_\text{sk} = \frac{1}{4\pi}\iint\limits_{\sigma} \bm{n}\cdot\left(\frac{\partial \bm{n}}{\partial x} \times \frac{\partial \bm{n}}{\partial y}\right)dxdy ~,
\]
where $\bm{n}(x,y)$ represents the Stokes vectors to construct a skyrmion and $\sigma$ denotes the skyrmion area. For the experimentally measured Stokes fields, we applied linear interpolation followed by a Gaussian filter as part of the post-processing procedure to suppress noise and enhance the accuracy of the skyrmion number calculation. Besides, due to the different size of the skyrmionic Stokes fields, careful area selection was crucial.

Ideally, the Stokes skyrmion is well-defined with clear circular symmetry, allowing the topological charge to be determined by integrating over a circular contour. However, turbulence disrupts this symmetry and a circular integral no longer gives us the correct topological charge. Therefore, it is necessary to accurately define an adaptive contour that accurately captures the topological charge. Here, we applied the Active Contour Model (ACM), otherwise known as Snakes\cite{kass1988snakes}. Snakes are an unparalleled tool widely employed in computer vision, primarily used to outline an object's boundary in a noisy two-dimensional environment. They are adaptive, energy-minimizing contours that evolve iteratively under the influence of internal and external forces, denoted by $E_{int}$ and $E_{ext}$, respectively. Mathematically, the Snake's energy function is given by:
\[
E_\text{snake}=\int_{0}^{1}(E_\text{int}+E_\text{ext}) \, ds ~,
\]
where $ds$ represents the differential element of the curve parameter. The optimization of the snake function is based on the concept of gradient descent. Starting from an initial contour, the algorithm iteratively minimizes the Snake function's energy, moving in the direction of the negative gradient at each step. This process continues until the local minima are found, at which point the contour outlines the boundary of the image. 

\clearpage
\newpage
\onecolumngrid \bigskip

\begin{center} {{\bf \large EXTENDED DATA}}\end{center}

\setcounter{figure}{0}
\makeatletter
\renewcommand{\thefigure}{E\@arabic\c@figure}

\begin{center}
	\begin{figure}[!ht]
		\includegraphics[width=1\linewidth]{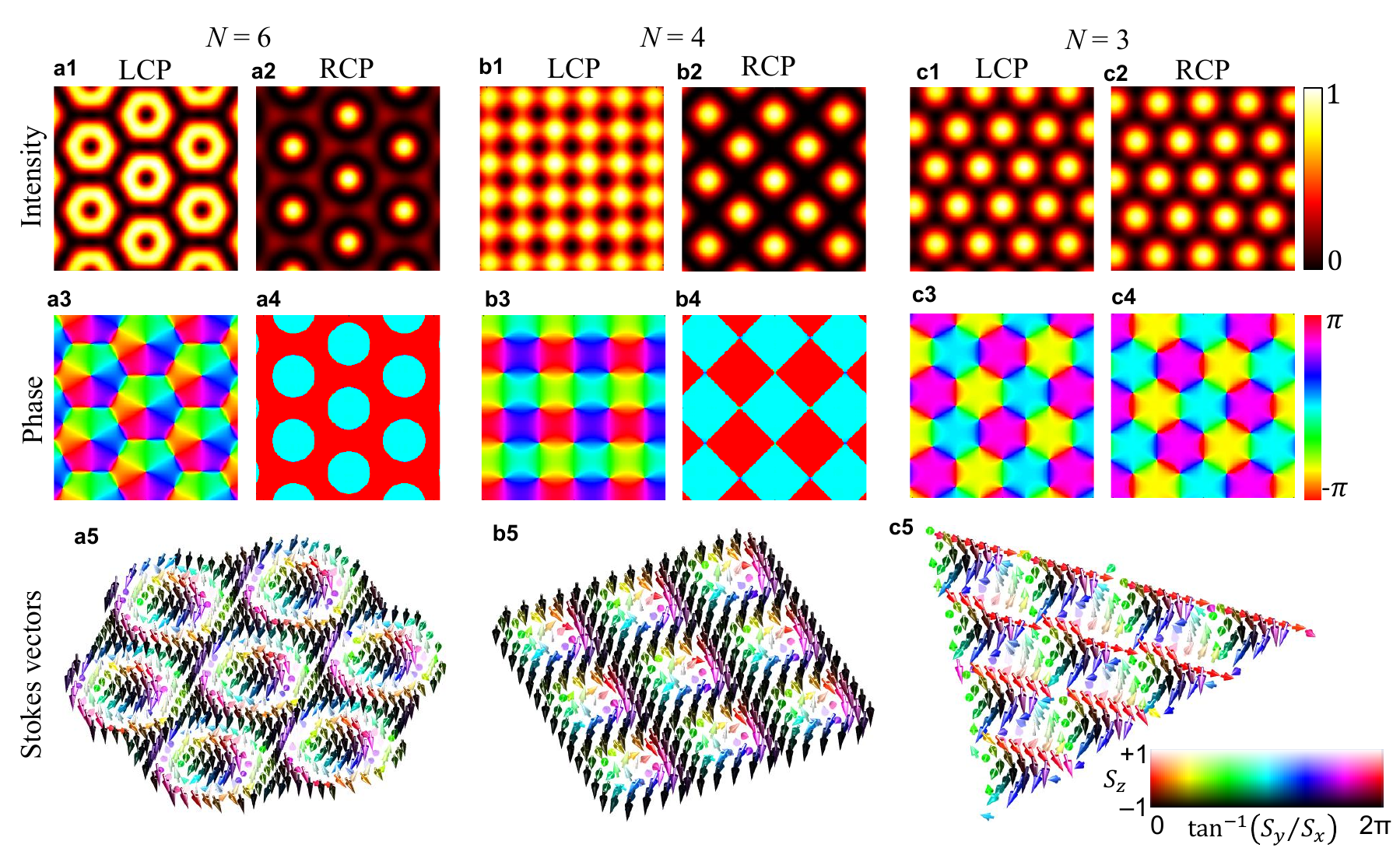}
		\caption{\textbf{Basic topological lattices.} (a1-a2) Intensity and (a3-a4) phase distribution of LCP and RCP whereas, (a5) stokes vector topological texture for C$_6$ skyrmionium lattices. (b1-b2) Intensity and (b3-b4) phase distribution of LCP and RCP whereas, (b5) stokes vector topological texture for C$_4$ skyrmion lattices. (c1-c2) Intensity and (c3-c4) phase distribution of LCP and RCP whereas, (c5) stokes vector topological texture for C$_3$ meron-antimeron lattices.  }
		\label{fig:FigE1}
	\end{figure}
\end{center}

\newpage 

\begin{center}
	\begin{figure}[!ht]
		\includegraphics[width=1\linewidth]{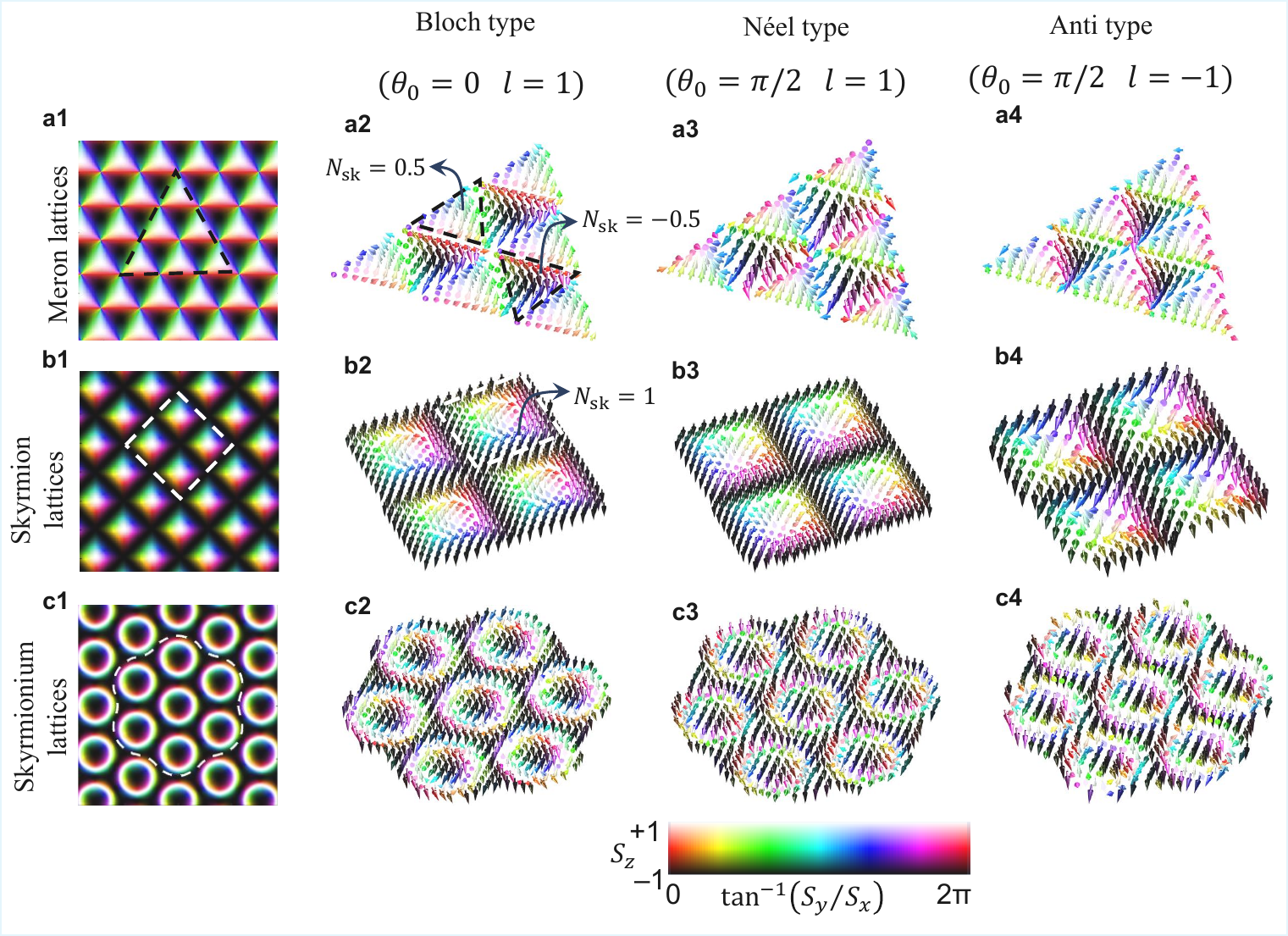}
		\caption{\textbf{Diverse topological lattices.} (a1-a4) Skyrmion texture and stokes vectors for Bloch, N\`eel and anti type C$_3$ meron-anti meron lattices. (b1-b4) Skyrmion texture and stokes vectors for Bloch, N\`eel and anti type C$_4$ skyrmion lattices. (c1-c4) Skyrmion texture and stokes vectors for Bloch, N\`eel and anti type C$_6$ skyrmionium lattices. }
		\label{fig:FigE2}
	\end{figure}
\end{center}

\newpage 

\begin{center}
	\begin{figure}[!ht]
		\includegraphics[width=1\linewidth]{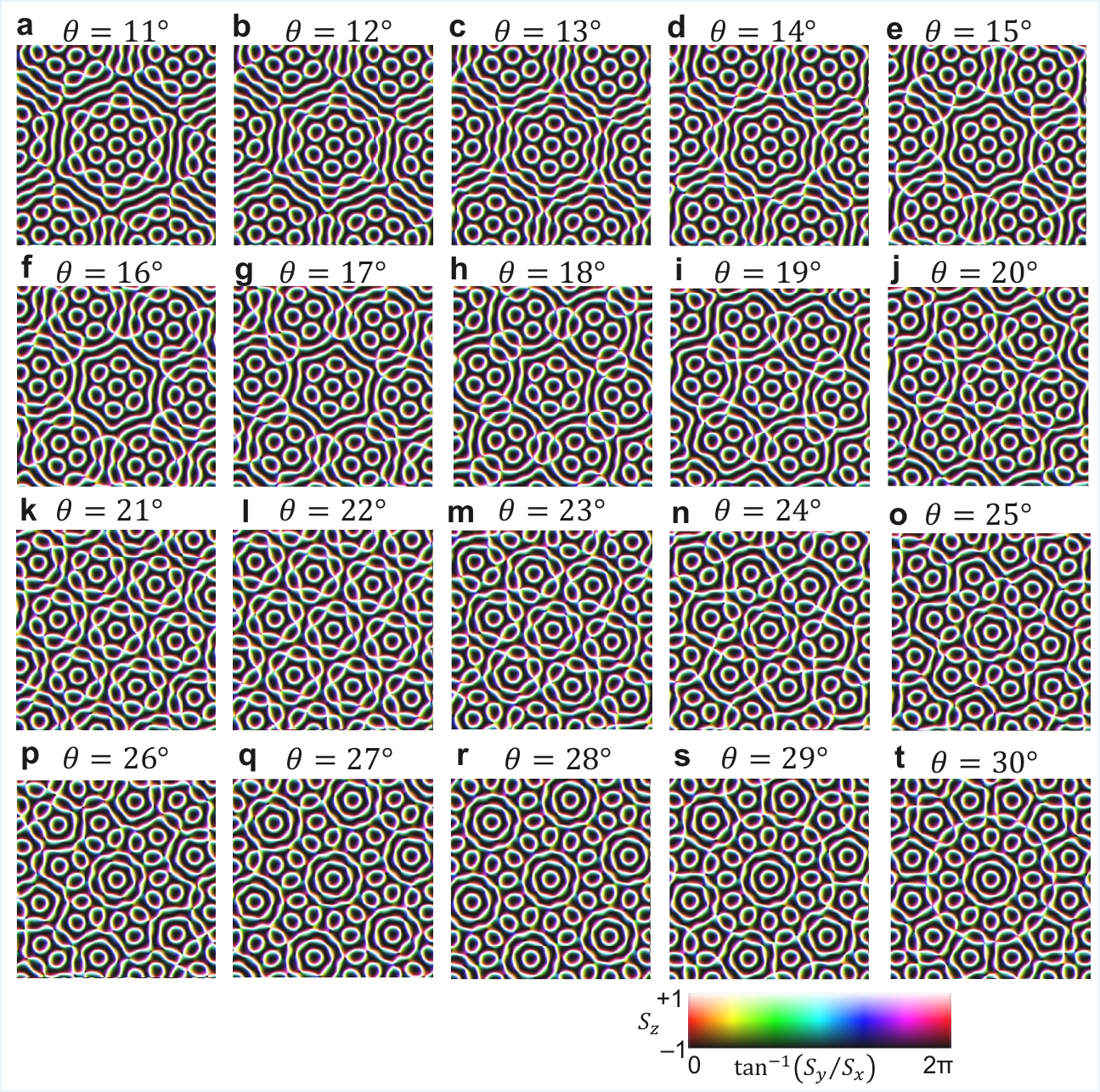}
		\caption{\textbf{Different supertopologies for various twist angles.} (a-t) skyrmion texture for supertopologies obtained for twist angles $\theta=11^\circ-30^\circ$ for C$_6$ skyrmionium lattices.}
		\label{fig:FigE3}
	\end{figure}
\end{center}

\newpage 

\begin{center}
	\begin{figure}[!ht]
		\includegraphics[width=1\linewidth]{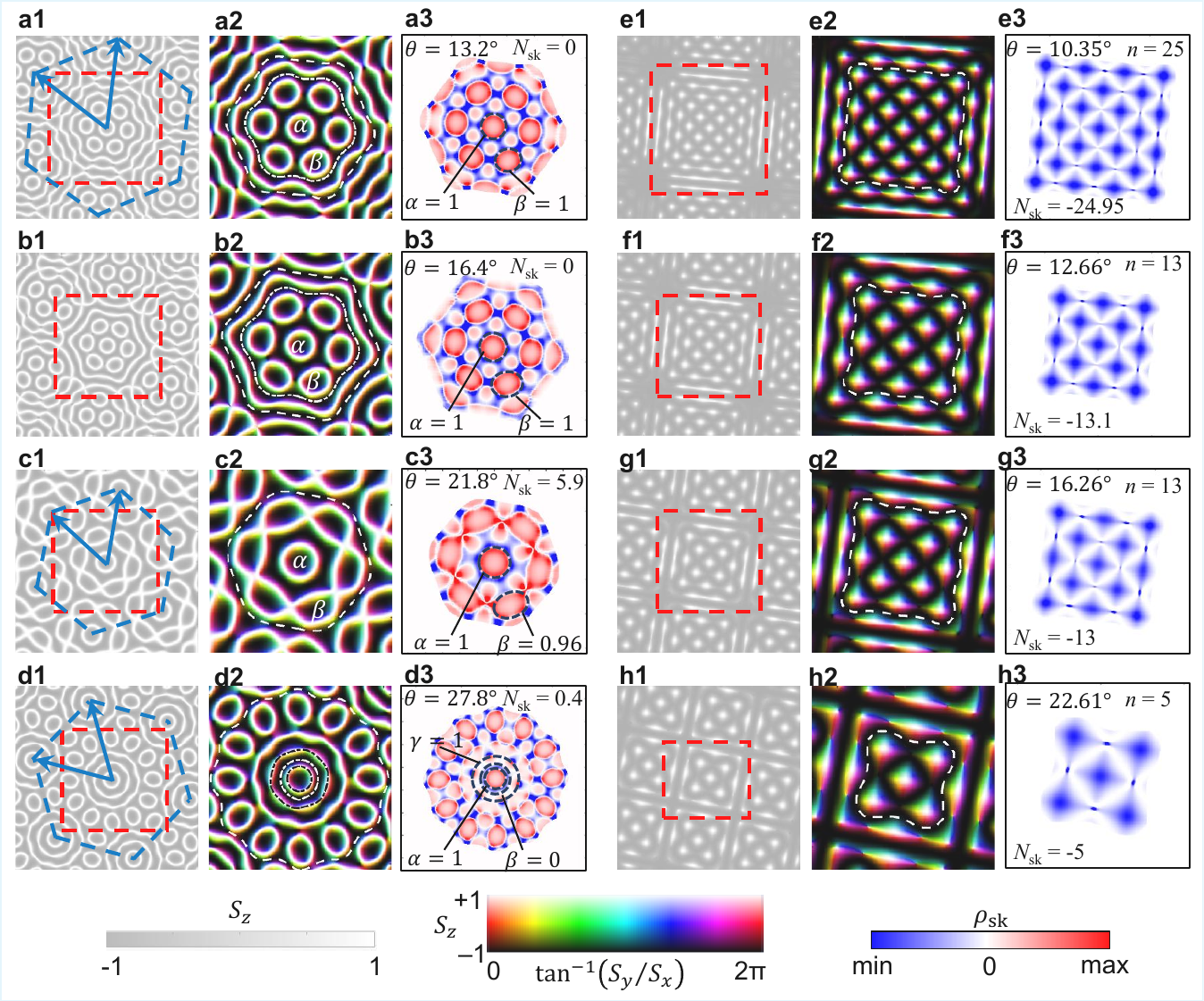}
		\caption{\textbf{Simulation results for various supertopologies.} }
		\label{fig:FigE4}
	\end{figure}
\end{center}

\newpage 


\begin{center}
	\begin{figure}[!ht]
		\includegraphics[width=1\linewidth]{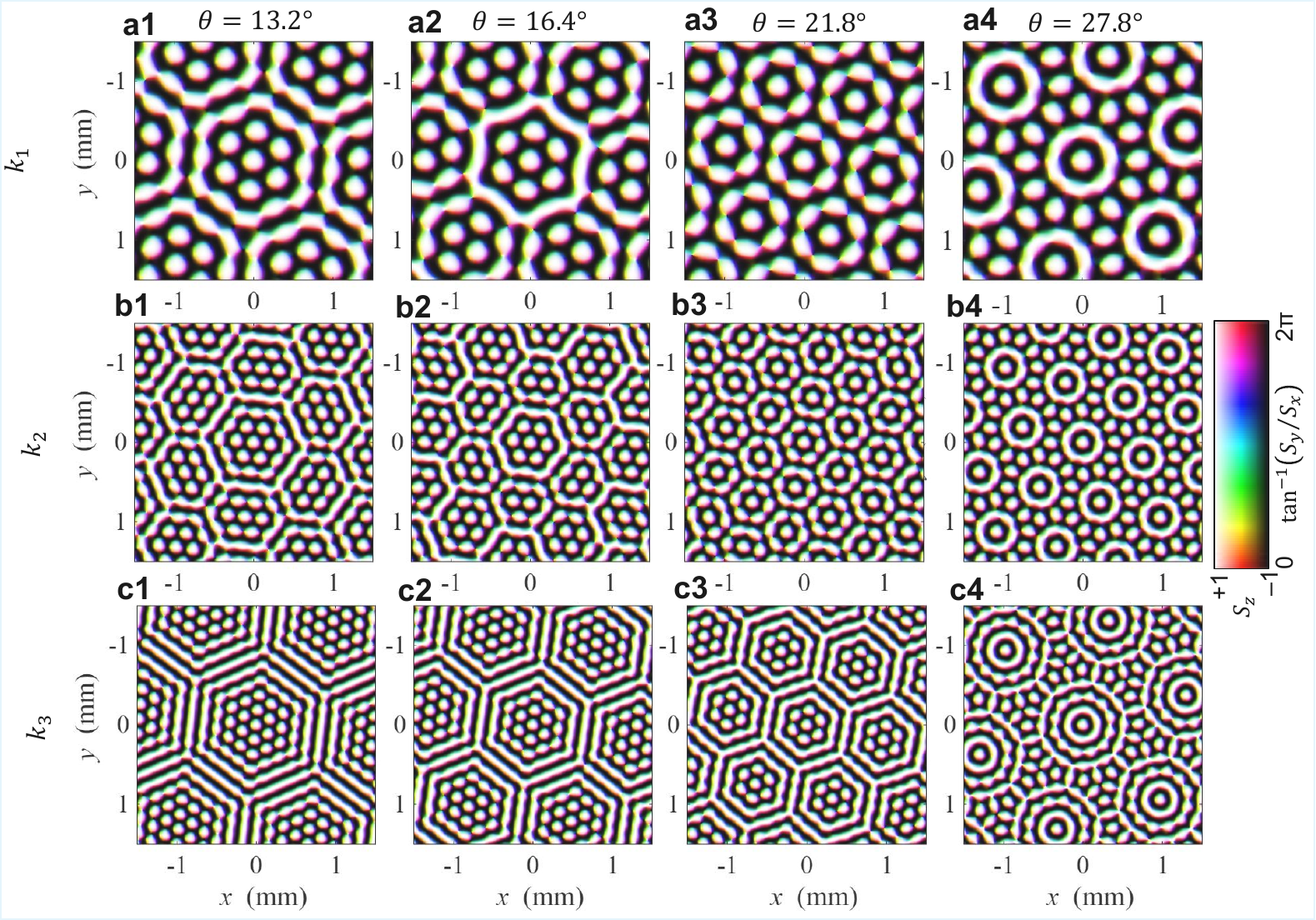}
		\caption{\textbf{Hierarchical supertopology decomposition.} }
		\label{fig:FigE5}
	\end{figure}
\end{center}

\begin{center}
	\begin{figure}[!ht]
		\includegraphics[width=1\linewidth]{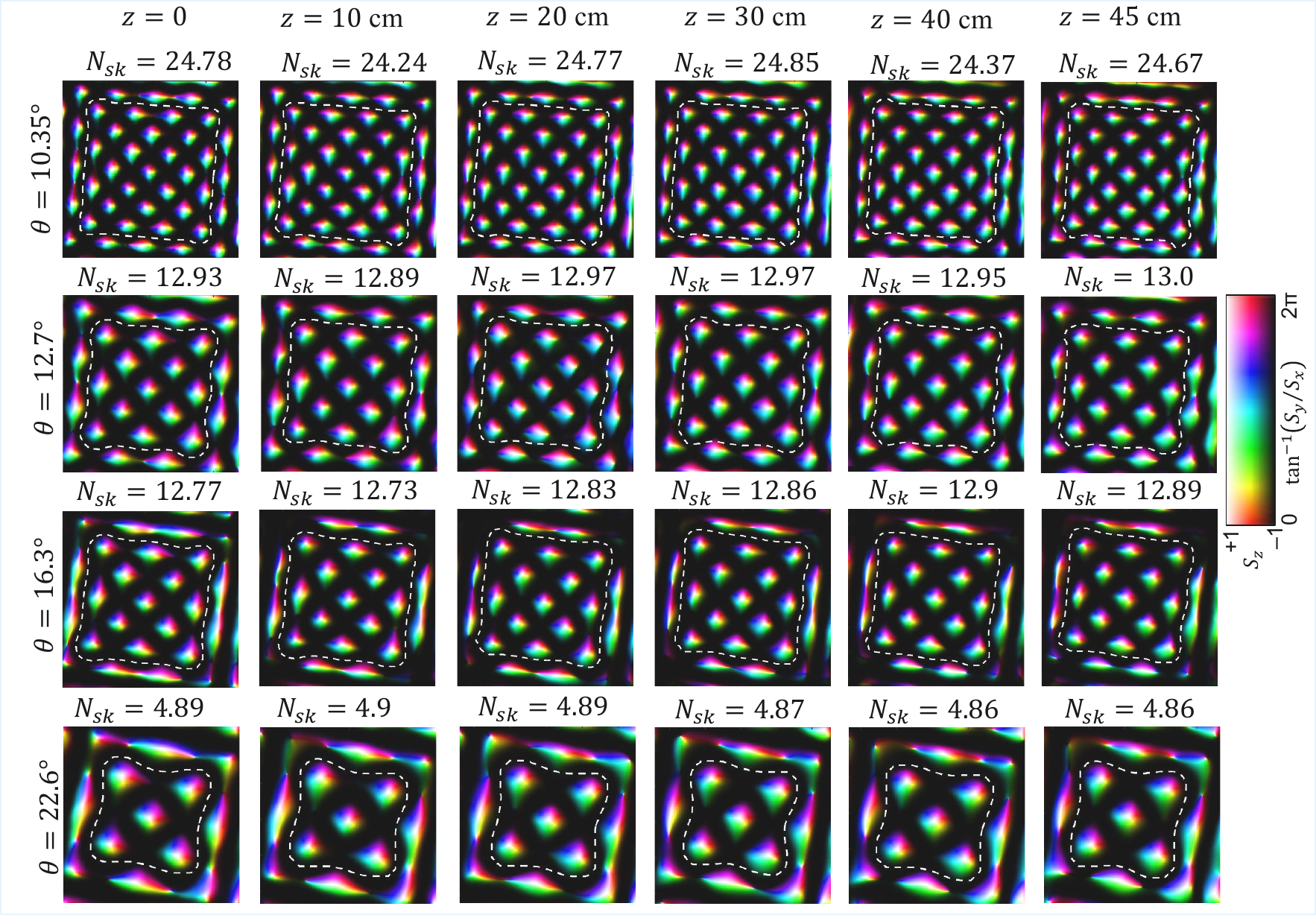}
		\caption{\textbf{Experimental results for stable propagation of C$_4$ skyrmion cluster super lattices.} }
		\label{fig:FigE6}
	\end{figure}
\end{center}

\begin{center}
	\begin{figure}[!ht]
		\includegraphics[width=1\linewidth]{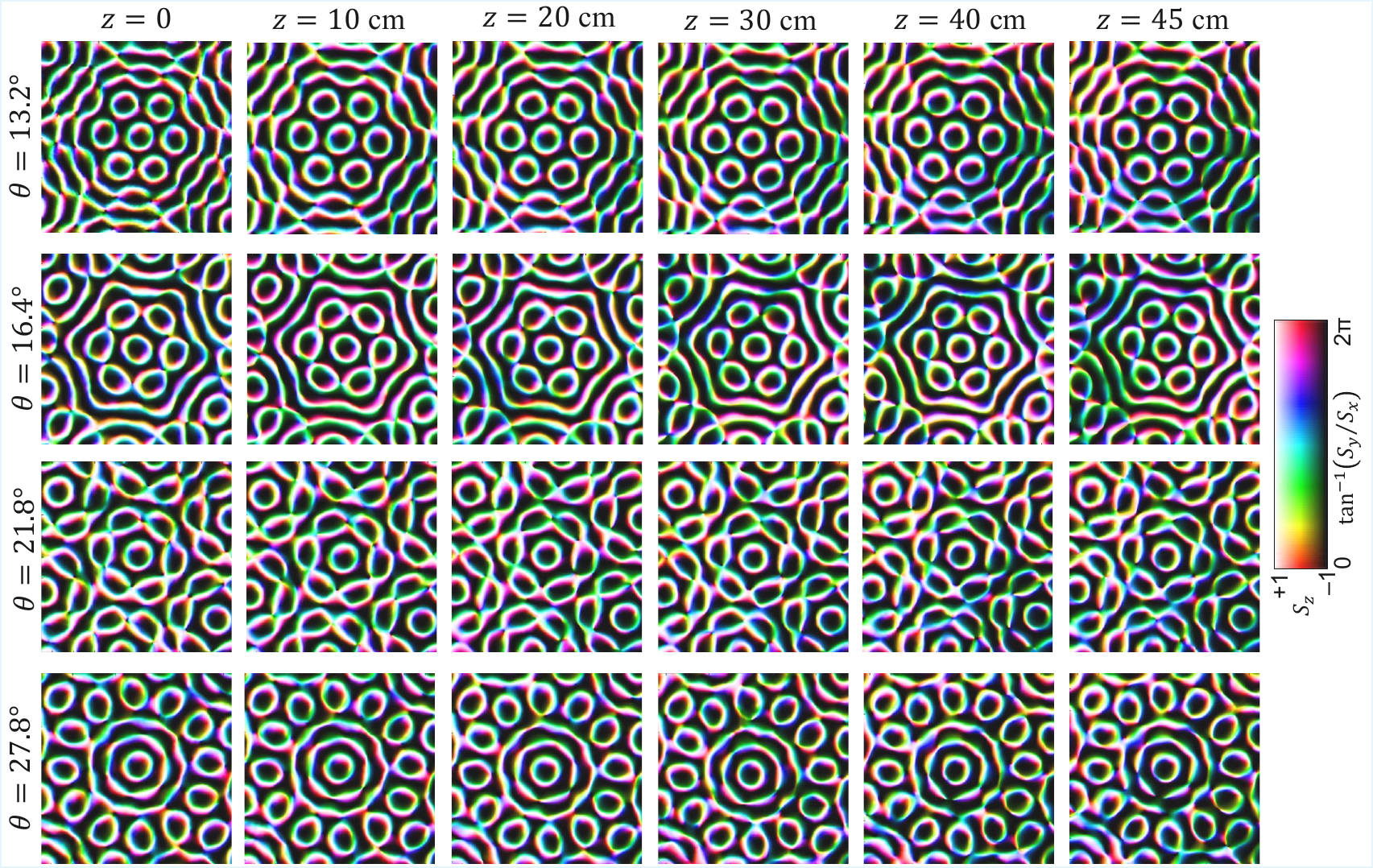}
		\caption{\textbf{Experimental results for stable propagation of C$_6$ super topologies.} }
		\label{fig:FigE7}
	\end{figure}
\end{center}

\begin{center}
	\begin{figure}[!ht]
		\includegraphics[width=1\linewidth]{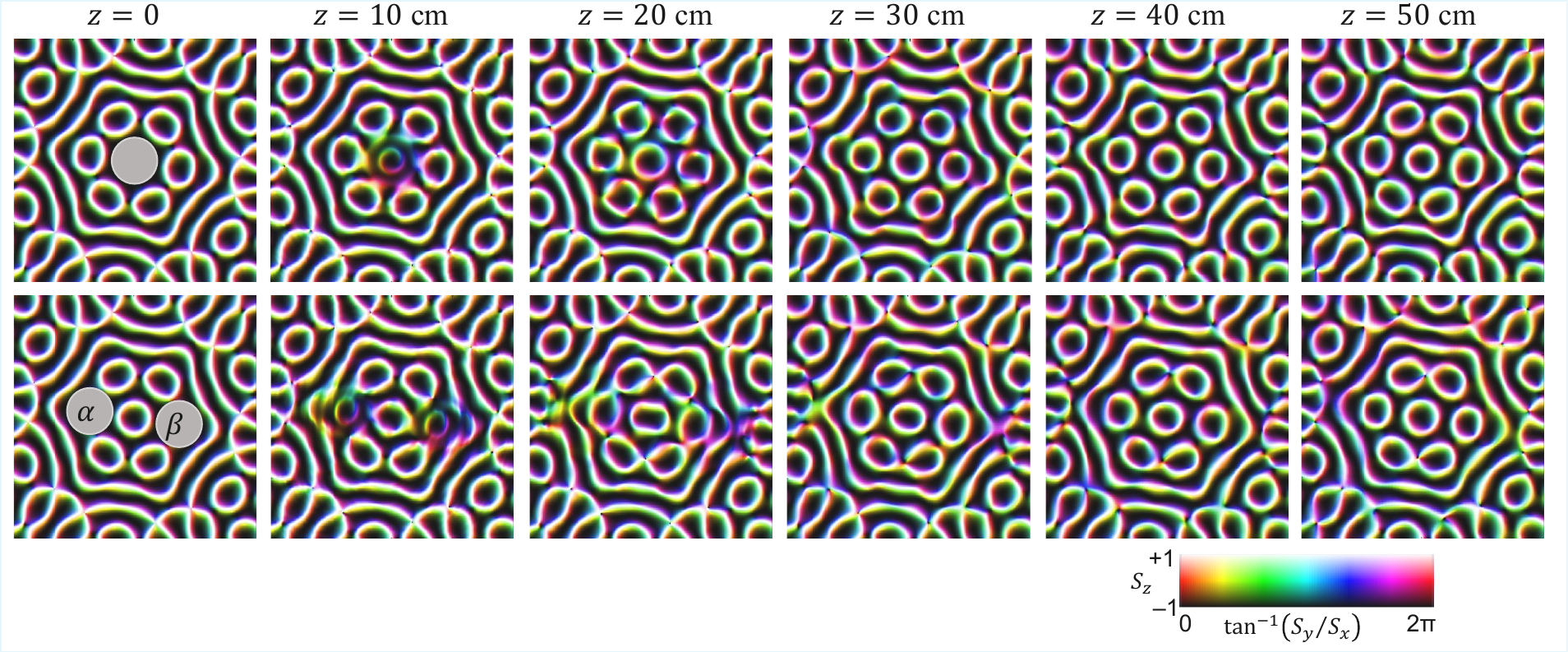}
		\caption{\textbf{Simulation results for self-healing.} }
		\label{fig:FigE8}
	\end{figure}
\end{center}

\end{document}